# Development of CODO: A Comprehensive Tool for COVID-19 Data Representation, Analysis, and Visualization

Biswanath Dutta[1] and Debanjali Bain[2]

[1, 2] Documentation Research and Training Centre (DRTC), Indian Statistical Institute, Bangalore, India
[1] bisu@drtc.isibang.ac.in

[2] Department of Library and Information Science, Calcutta University, Kolkata, India
[2] debanjali@drtc.isibang.ac.in

**Abstract:** Artificial intelligence (AI) has become indispensable for managing and processing the vast amounts of data generated during the COVID-19 pandemic. Ontology, which formalizes knowledge within a domain using standardized vocabularies and relationships, plays a crucial role in AI by enabling automated reasoning, data integration, semantic interoperability, and extracting meaningful insights from extensive datasets. The diversity of COVID-19 datasets poses challenges in comprehending this information for both human and machines. Existing COVID-19 ontologies are designed to address specific aspects of the pandemic but lack comprehensive coverage across all essential dimensions. To address this gap, CODO, an integrated ontological model has been developed encompassing critical facets of COVID-19 information such as aetiology, epidemiology, transmission, pathogenesis, diagnosis, prevention, genomics, therapeutic safety, and more. This paper reviews CODO since its inception in 2020, detailing its developments and highlighting CODO as a tool for the aggregation, representation, analysis, and visualization of diverse COVID-19 data. The major contribution of this paper is to provide a summary of CODO's development, and outline the overall development and evaluation approach. By adhering to best practices and leveraging W3C standards, CODO ensures data integration and semantic interoperability, supporting effective navigation of COVID-19 complexities across various domains.

**Keywords:** COVID-19, Coronavirus, Ontology, CODO, Model, Analytics, Methodology, Reasoning

## 1. Introduction

The surge of COVID-19 datasets encompasses diverse aspects including virology, symptoms, treatments, genomic sequences, and patient information (Dhama et al., 2020). This heterogeneous data presents challenges in making COVID-19 information comprehensible to both humans and machines, hampering comprehensive understanding and effective utilization for pandemic-related queries (Filip et al., 2022; Jean et al., 2020). Ontology has emerged as a pivotal approach for integrating and comprehending massive COVID-19 data, facilitating data integration, semantic interoperability, and advanced retrieval (Dutta, Nandini, et al., 2015). Various types and sources of datasets are available for COVID-19 information, such as websites, guidelines, gene banks, literature, and graphical data, each following heterogeneous formats (Dutta et al., 2022). Ontology-based approaches, exemplified by frameworks like CIDO (He et al., 2020, 2022), COVOC (Caucheteur et al., 2023), COVID-CRF-RAPID (Bonino, 2020), CODO (Dutta & DeBellis, 2020), ROC (Qundus et al., 2021), and CIDO-COVID-19 (Xiao et al., 2021), are gaining traction. However, our previous research highlighted that no single ontology comprehensively covers all essential dimensions of COVID-19 specific information or dataset (Bain & Dutta, 2023). Existing COVID-19 ontologies focus on specific aspects of the pandemic, lacking comprehensive coverage across critical dimensions. To address this gap, we have developed the integrated CODO model, encompassing key facets of COVID-19 information such as aetiology, epidemiology, transmission, pathogenesis, diagnosis, prevention, genomics, and therapeutic safety. This paper provides a detailed review of recent updates to the COVID-19 Ontology (CODO). The objective is to provide recent developments and updates about CODO, serving as an ontological framework for aggregating and analysing diverse COVID-19 data, integrating epidemiological, clinical, and genomic information. CODO is patient-centric and has been developed with the intent to integrate patient-related data effectively. Version 1.4, known as COVIDRO, addresses treatment challenges, patient risk factors, and drug interactions. Version 1.5, or COVID-19 Virus Genomics Ontology (CODO_VGO), enhances genomic data representation, incorporating information on variants, mutations, genes, and proteins. These updates enhance CODO's capabilities for advanced analytics, supporting stakeholders such as ontology developers, medical professionals and policymakers. The major contribution of this paper is to provide a summary and description of each CODO version's development, alongside outlining the methodology and evaluation criteria for building CODO. Adhering to best practices and leveraging W3C standards, CODO ensures data integration and semantic interoperability, making it an essential tool for navigating COVID-19 data complexities across various domains.

The paper is organized as follows: Section 2 delves into the related work. Section 3 outlines the CODO development approach, detailing its evolution since 2020. Section 4 presents the results and discussion, describing the CODO ontology and its modules, and examining the integration and applications of CODO. Finally, Section 5 concludes the paper and discusses future work.

## 2. Related Work

Significant efforts have been made to develop ontologies that represent and organize knowledge and data related to COVID-19. These ontologies aim to structure and facilitate the use of the vast amount of information generated during the pandemic. Key contributions in this field include the COVID-19 Surveillance Ontology (de Lusignan et al., 2020), CIDO-COVID-19 (Xiao et al., 2021), COVIDCRFRAPID (Bonino, 2020), DRUGS4COVID19 (Badenes-Olmedo et al., 2020), ROC (Qundus et al., 2021), and CODO (Dutta & DeBellis, 2020), among others. COVID-19 Surveillance Ontology is designed to track COVID-19 cases and related respiratory illnesses. It utilizes data from electronic medical record systems to monitor the spread and impact of the virus, ensuring timely and accurate surveillance. Adhering to recommended ontology development criteria, CIDO-COVID-19 covers a broad range of aspects related to the disease. These include disease characteristics, diagnosis, transmission mechanisms, symptoms, therapeutic interventions, and preventive measures. This comprehensive approach ensures a representation of COVID-19-related knowledge. Developed by the World Health Organization (WHO), COVIDCRFRAPID ontology serves as a data model for representing the WHO's COVID-19 RAPID case record forms. It provides a structured framework for capturing and querying detailed information from these forms, facilitating standardized data collection and analysis. DRUGS4COVID19 ontology details the relationships between medications and the COVID-19 virus. It includes categories such as drugs, their effects, related diseases, symptoms, and chemical substances, thus supporting research and development in the pharmacological domain. The ROC ontology evaluates governmental responses to the pandemic, focusing on both positive and adverse outcomes. This evaluation helps in understanding the effectiveness of various measures implemented to control the spread of the virus and mitigate its impact. Despite these significant contributions, a thorough review of 24 COVID-19 ontologies has revealed gaps in comprehensive coverage across various domains and dimensions of the pandemic (Bain & Dutta, 2023). These gaps underscore the need for an integrated model that can provide a more holistic view of COVID-19 data. In response to this need, our study introduces an integrated ontology model. This model aims to address the identified gaps by consolidating diverse COVID-19-related data into a unified framework. For a detailed discussion on specific dimensions and further analysis, readers are referred to our previous work (Bain & Dutta, 2023) and comprehensive studies on this topic.

## 3. Methodology

This section outlines the overall development approach and evaluation of our COVID-19 Ontology (CODO). Before delving into the methodology, we first discuss the evolution of CODO from its inception in 2020 to the present day.

### 3.1 Evolution of CODO: From 2020 to Present

The development of CODO began in response to the emerging COVID-19 pandemic. As the severity of the outbreak became evident in early 2020, the need for a structured framework to manage the vast and complex data related to the virus was clear. The first author envisioned this necessity and, in collaboration with another researcher, produced the initial version of CODO 1.0[1], which was detailed in an early publication (Dutta & DeBellis, 2020). Initially, the focus was on understanding the immediate data needs and complexities associated with COVID-19. As the pandemic progressed, the volume and complexity of datasets published increased significantly, prompting further development of the ontology. Early datasets included epidemiological data, clinical records, resource availability, genomic sequences, and public health information, revealing the multifaceted nature of COVID-19 data (Dutta et al., 2022). The initial development phase of CODO was iterative, involving continuous brainstorming, discussions, and refinements. The authors identified the need to expand the ontology's breadth and depth across various domains and facets of the pandemic. This iterative process led to the identification of gaps in existing ontologies and the need for a more comprehensive framework. One significant milestone in CODO's evolution was the publication of a paper identifying various domains and coverage areas specific to COVID-19 ontologies, highlighting existing lacunae and proposing a developmental approach (Bain & Dutta, 2023). This work laid the groundwork for the subsequent development of various dimensions and versions of CODO. For example, COVIDRO (COVID-19 Risk Ontology) was developed to address treatment challenges, patient risk factors, and drug interactions. Another significant version, COVID-19 Virus Genomics Ontology (CODO_VGO), enhanced the representation of genomic data, incorporating information on variants, mutations, genes, and proteins. The development of CODO has been an ongoing process, characterized by iterative improvements and expansions. Each version built upon the knowledge gained from previous iterations, incorporating new data and insights. The process involved extensive domain knowledge in virology, aetiology, and pathogenesis related to COVID-19. Currently, the development of CODO continues as the authors strive to identify new directions and accommodate unstructured COVID-19-specific information within the model. The subsequent sections provide further details on the development and evaluation of CODO.

### 3.2 CODO Development Approach

There are various ontology development methodologies available, which include, METHONTOLOGY (Fernández-López et al., 1997), DILIGENT (Vrandečić et al., 2005), NeOn (Suárez-Figueroa et al., 2012), YAMO (Dutta, Chatterjee, et al., 2015), etc. The

---

[1] Available from https://w3id.org/codo/1.0

development of the CODO ontology follows a systematic and structured approach inspired by established methodologies in ontology design, YAMO and NeOn. The CODO development process is comprehensive, involving several distinct phases that ensure a formal, modular and semantically rich ontology (DeBellis & Dutta, 2022). While this paper does not delve into the detailed methodology for CODO's development, readers can refer to dedicated articles for individual CODO modules. Here is a brief overview of the general CODO development methodology (refer to Fig. 1):

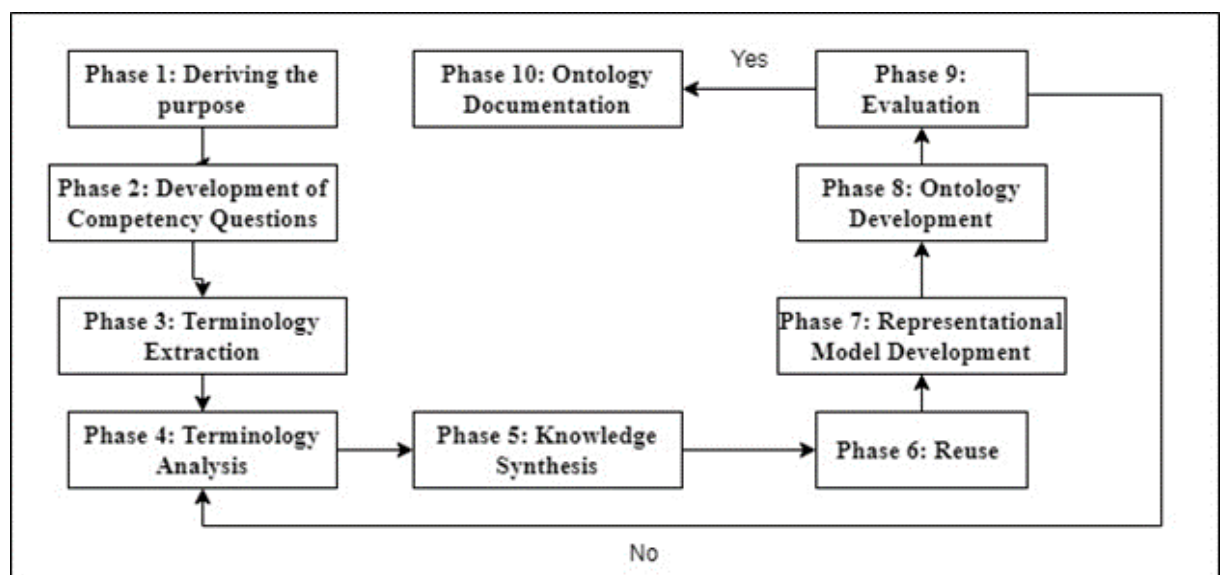

**Figure 1:** Overview of CODO Development Approach

*Phase 1: Define the Purpose:* The initial step involves defining the purpose and scope of the CODO ontology. This includes identifying the primary objectives and potential use cases. The CODO ontology aims to provide a comprehensive framework for representing and analyzing COVID-19-related information. It is designed to support various stakeholders, including government agencies, healthcare institutions, researchers, and data publishers, in annotating and describing information effectively.

*Phase 2: Development of Competency Questions:* Competency questions are developed to expand on the purpose defined in Phase 1. These questions guide the ontology development, ensuring it encompasses the necessary information and relationships. They focus on critical aspects such as risk factors, mutations, variants, therapeutics, drug adverse effects, drug interactions, etc. This phase ensures the ontology can meet specific informational needs and support informed decision-making. Examples of competency questions include: What mutations were discovered in patient p1's genomic testing results? Which SARS-CoV-2 variant has patient p1 been diagnosed with? Which therapeutics are recommended for COVID-19 patients with specific underlying health conditions? What are the risk factors associated with severe COVID-19 infection? What are the potential adverse effects of specific COVID-19 drugs? Identify the drug interactions that may occur with a particular COVID-19 therapeutic. Which drugs should be avoided due to known interactions with commonly prescribed COVID-19 medications? How does weather condition affect the transmission of COVID-19? Provide the contact tracing details for patient p. These competency questions ensure that the ontology can address specific informational needs and support informed decision-making across various aspects related to COVID-19.

*Phase 3: Terminology Extraction:* This phase involves systematically extracting terminology related to the domain of interest. Authoritative sources, including medical guidelines, clinical trials, and literature, are reviewed to identify relevant terms. This process, also known as knowledge extraction, ensures a comprehensive set of terms is collected to form the foundation of the ontology. The extracted terms include concepts, definitions, and relationships pertinent to COVID-19 and its treatment options.

*Phase 4: Terminology Analysis:* The extracted terms are analyzed to determine their appropriate classification within the ontology. Complex and compound concepts are broken down into fundamental entities. Each term is examined to decide whether it should be categorized as a class, property, or another relevant type, based on its inherent definition and properties. This analysis ensures accurate representation and organization within the ontology.

*Phase 5: Knowledge Synthesis:* In this phase, the relationships between the extracted concepts are established, and a comprehensive class hierarchy is synthesized. The concepts are organized into a structured hierarchy, capturing their semantic relationships and properties. This structured organization facilitates systematic representation and navigation of the domain knowledge within the ontology.

*Phase 6: Reuse of Existing Ontologies:* To promote data consistency and enhance compatibility, the CODO ontology leverages pre-existing vocabularies and ontologies. Well-known vocabularies like Schema.org, FOAF[2], ORG[3], and SNOMED-CT[4] are reused where applicable. This practice ensures interoperability and facilitates integration with other datasets that utilize the same vocabulary.

*Phase 7: Representational Model Development:* A representational model is developed to effectively capture and organize the domain knowledge. This model defines the various classes, properties, and their relationships within the ontology. It provides an illustrative framework that showcases the structure and components of the ontology, ensuring comprehensive coverage and coherence.

*Phase 8: Ontology Development:* The ontology is developed using OWL-DL, a W3C recommended description logics-based ontology language. Tools such as the Protégé ontology editor (*Protégé*, n.d.), along with plugins like Pellet Reasoner (Sirin et al., 2007) and SPARQL (DuCharme, 2013), are utilized to build the ontology. This phase involves formalizing the ontology components, including concepts, properties, and hierarchical relationships.

*Phase 9: Evaluation*: The ontology undergoes a rigorous evaluation to ensure consistency, completeness, and effectiveness. Tools like Pellet reasoner[5], OntoDebug[6], OOPS![7], and SPARQL[8] are employed to validate the ontology's quality. This evaluation process helps identify potential improvements and ensures the ontology meets the desired objectives.

---

[2] http://xmlns.com/foaf/0.1/
[3] https://schema.org/
[4] https://www.snomed.org/
[5] https://www.w3.org/2001/sw/wiki/Pellet
[6] https://protegewiki.stanford.edu/wiki/OntoDebug
[7] https://oops.linkeddata.es/
[8] https://www.w3.org/TR/sparql11-query/

*Phase 10: Ontology Documentation*: Comprehensive documentation is created to facilitate understanding, dissemination, and reuse of the ontology. Tools like WIDOCO[9] (Wizard for Documenting Ontologies) are used to generate detailed documentation. This includes information on classes, properties, individuals, interactive visualization of the ontology, and ontology metrics, presented in a user-friendly web interface. The documentation also provides insights into the ontology's structure, usage, and versioning information. The CODO development approach ensures a systematic and thorough process for creating a formal ontology. By following established methodologies and leveraging best practices, the CODO ontology effectively supports the representation and analysis of complex domain knowledge, particularly in the context of COVID-19.

## 4. Results and Discussion

CODO (https://w3id.org/codo) is an ontological framework specifically designed for the aggregation and analysis of COVID-19 data. The latest update, integrated CODO version 1.5, significantly enhances the ontology with a comprehensive set of terms and relationships tailored to capture a broad spectrum of COVID-19-related information. The extensive structure of CODO 1.5, comprising 385 classes, 214 object properties, and 123 data properties, makes it an indispensable tool for addressing the challenges posed by the COVID-19 pandemic across various domains and disciplines. This includes epidemiological data, clinical findings, etiology, diagnosis, therapeutic facilities, comorbidities, drug interactions, adverse effects, COVID-19 therapeutics, patient travel histories, symptoms, test results, and extensive genomic data. The ontology is designed to support applications such as search engines, question-answering systems, risk detection tools, document annotation tools, data knowledge graphs, and analytics tools, proving invaluable in navigating and interpreting vast amounts of COVID-19 data. In the subsequent paragraphs, we discuss some of the modules of CODO.

### 4.1 CODO modules

*4.1.1 Module for Collection, Representation, and Analysis of COVID-19 Pandemic Data*

This CODO module is dedicated to the collection, representation, and analysis of COVID-19 pandemic data, focusing on two main dimensions: detailed data on COVID-19 cases and patient-oriented information. The first dimension includes data on active, recovered, deceased, and migrated cases, as well as resource requirements and availability (e.g., ICU beds, invasive ventilators). This data is categorized by geolocation (district, state/province, country) and time (date, time). The second dimension covers patient information such as symptoms, severity levels, travel history, suspected reasons for contracting the disease, inter-patient relationships, daily diagnosis results, and COVID-19 medical facilities. This model supports semantic querying and retrieval, enabling advanced analytics like trend analysis and growth projections. Additionally, representing anonymous patient data and their relationships aids in analyzing disease behavior and transmission routes. This is the first CODO module, published as CODO 1.0[10].

*4.1.2 WEAR Module: WEather, Government Responses and measures Information Related to COVID-19[11]*

---

[9] https://github.com/dgarijo/Widoco
[10] https://w3id.org/codo/1.0
[11] https://w3id.org/codo/1.3

The WEAR module facilitates the integration of weather-specific information, government responses, and public health measures related to COVID-19 (refer to Fig. 2 for a snippet). By correlating weather conditions (e.g., temperature, humidity, and wind speed) with the spread and severity of COVID-19, this module helps derive valuable insights into how environmental factors and public health measures influence virus transmission, spread patterns, and more. The module includes classes such as WeatherCondition, Temperature, Humidity, WindSpeed, and Precipitation, with properties like hasTemperature, hasHumidity, hasWindSpeed, and hasPrecipitation. Specific instances include conditions like "HighHumidity" and "LowTemperature." The WEAR module also captures public health responses and social distancing measures, providing a comprehensive view of their impact on the COVID-19 situation. This includes classes and properties such as PublicHealthMeasure, SocialDistancing, ImposedPreventiveMeasure, responseToCOVID-19, and GovernanceAndSocio-economicMeasures. The WEAR module aims to bring together all aspects of seasonality patterns and public health responses for explicit data representation and analysis. It helps in understanding how lockdowns, vaccinations, and seasonal variations influence COVID-19 case numbers. For example, the module can analyze whether the implementation of lockdowns resulted in a decrease in cases, how vaccination timing affected case numbers, and the impact of seasonal changes on the virus's spread.

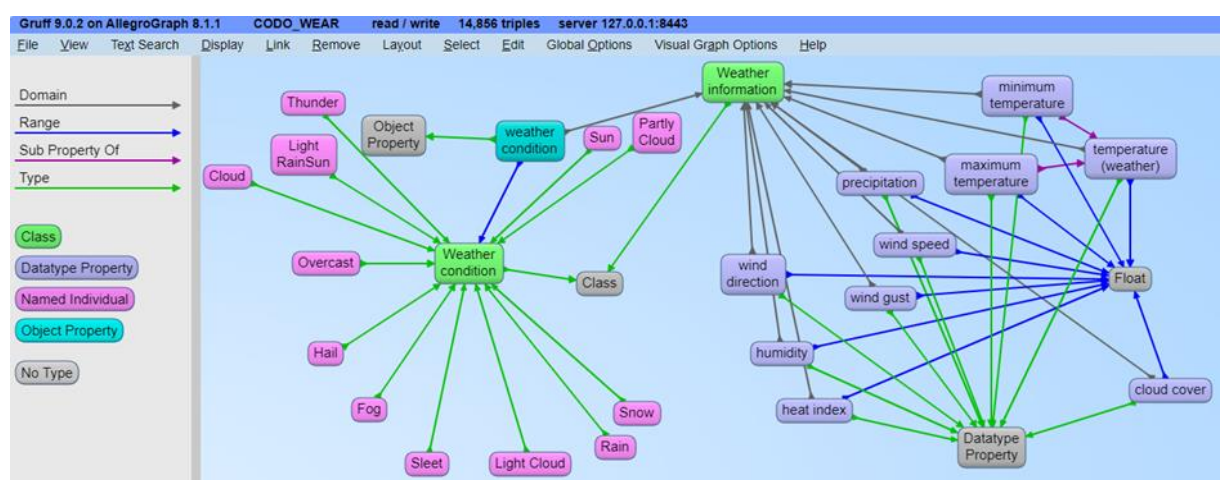

**Figure 2**: Ontological Representation of Weather Information Related to COVID-19 in Allegro's Gruff Visualization Tool

*4.1.3 COVIDRO Module[12]: Disease Drug Adverse Effects, Interactions, and Risk Factors*

The COVIDRO module captures data related to drug interactions, adverse effects, and risk factors associated with COVID-19 treatments (refer to Fig. 3). It includes detailed information on various medications used in the treatment of COVID-19, documenting their potential interactions and side effects. This module also considers patient-specific factors such as comorbidities and ongoing medications, providing a comprehensive view of potential risks. For example, it highlights interactions between antiviral drugs like Remdesivir and common medications such as anticoagulants, helping clinicians make informed decisions about patient treatment plans. The COVIDRO module includes classes like AdverseEffect, DrugInteraction, RiskFactor, Medication, Patient, and Symptom, and properties such as hasAdverseEffect, interactsWith, hasRiskFactor, and experiences, to capture data related to drug interactions (e.g.,

---

[12] https://w3id.org/codo/1.4

Ritonavir-BoostedNirmatrelvir/Carbamazepine DDI), adverse effects (e.g., Headache, Nausea), and risk factors (e.g., Age, ComorbidCondition) associated with COVID-19.

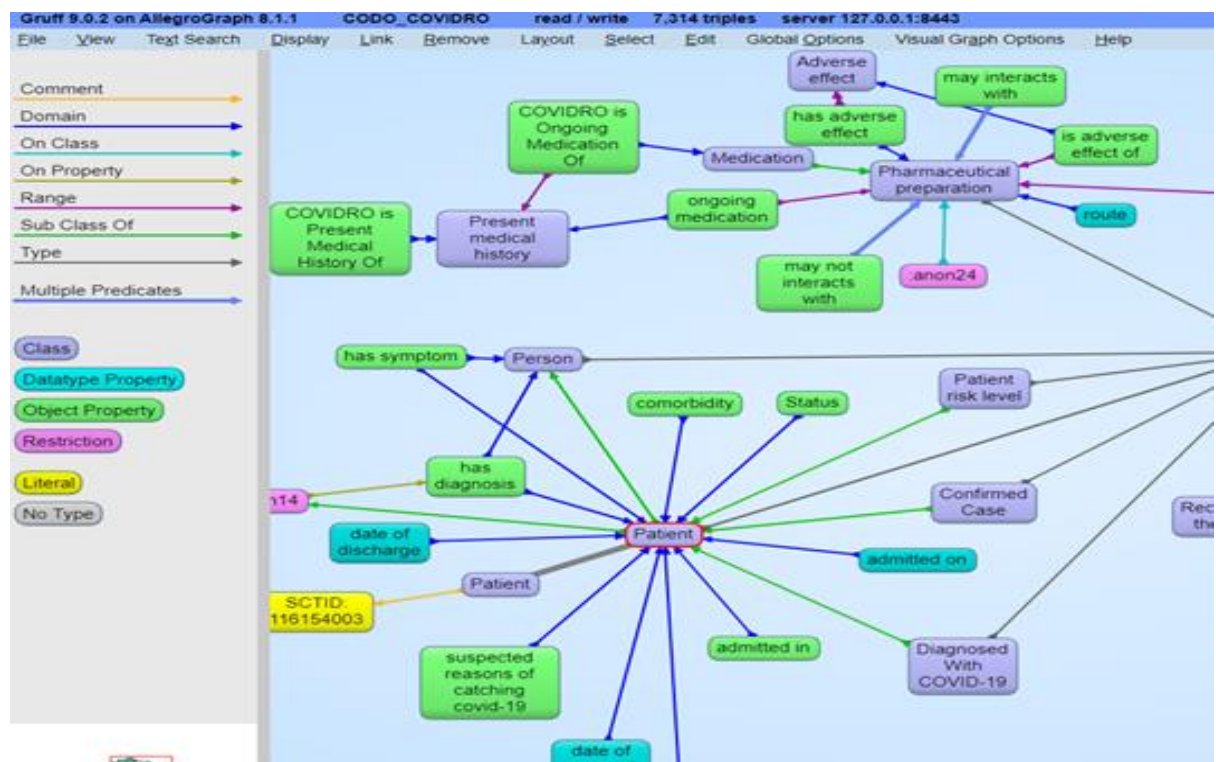

**Figure 3**: Ontological Representation of COVID-19 Disease Drug Adverse Effects, Interactions, and Risk Factors in Allegro's Gruff Visualization Tool

*4.1.4 VGO Module[13]: Variants, Genes, Mutations, and Genome Sequencing Data*

The VGO module is dedicated to genomic information, covering sequences, variants, mutations, and the methodologies involved in genome sequencing (see, Fig. 4). This module is essential for tracking the evolution of SARS-CoV-2 and understanding the genetic changes that affect virus transmissibility and vaccine efficacy. By integrating data from global repositories such as GISAID, the VGO module enables researchers to monitor the emergence of new variants and their geographic spread. It provides detailed genomic information, including specific mutations within the virus, which helps scientists correlate genetic changes with clinical outcomes and epidemiological trends. The VGO module includes classes such as Variant, Gene, Mutation, GenomeSequencing, Sample, and SequencingTechnology, and properties like hasGene, hasMutation, sequencedUsing, and derivedFrom, facilitating comprehensive genomic analysis.

---

[13] https://w3id.org/codo/1.5

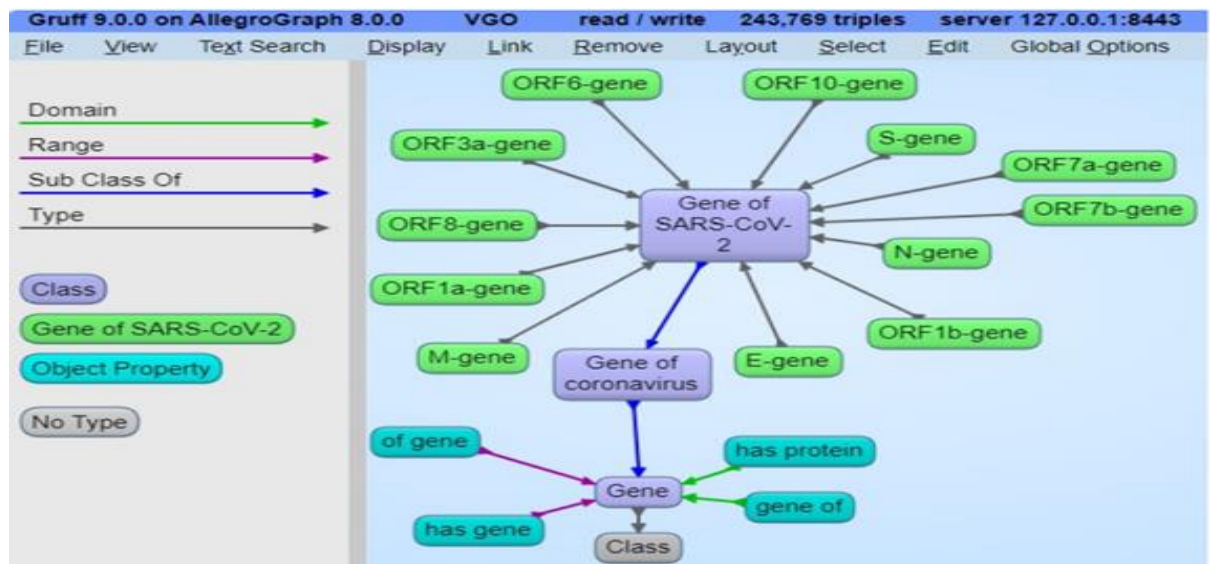

**Figure 4**: Ontological Representation of Variants, Genes, Mutations, and Genome Sequencing Data in Allegro's Gruff Visualization Tool

## 4.2 CODO, as an Integrated Model and its Applications

In this current work, the approaches for integration and aggregation of various modules are not within the scope and will be addressed in our ongoing future work. To ensure independence from external ontologies, we have established a dedicated namespace located at https://w3id.org/codo. This namespace functions as a permanent URL service, redirecting to https://www.isibang.ac.in/ns/codo/index.html. CODO is structured as a modular ontology, where each module is identified using URLs formatted as https://w3id.org/codo#ModuleName, such as https://w3id.org/codo#COVIDRO. This design facilitates seamless integration and expansion of modules like COViDRO within the overarching CODO ontology. Each concept within a specific module is uniquely identified using URIs structured as https://w3id.org/codo#ModuleName_ConceptName. For example, a concept class within COViDRO, like "Adverse effect", is represented as https://w3id.org/codo#COVIDRO_AdverseEffect. This standardized approach ensures a well-structured framework for concepts within CODO, promoting interoperability and facilitating their use across various applications and domains. Therefore, when a user implements a specific module within a system, it remains usable and integrates effectively with the broader CODO ontology**.** By providing a structured and detailed framework, CODO can enable the development of various applications that streamline and enhance the management of COVID-19 data.

These applications encompass:
   i. *Search Engines: To find specific information related to COVID-19 quickly and efficiently.*
   ii. *Question-Answering Systems: To delineate infection pathways and detect risk factors.*
   iii. *Risk Detection Tools: To identify and manage potential COVID-19 risks in populations.*
   iv. *Document Annotation Tools: To assist in the classification and analysis of COVID-19 related literature.*
   v. *Data Knowledge Graphs: To visualize relationships and connections between different COVID-19 data points.*

> vi. *Analytics Tools: To perform complex data analysis and derive insights from large datasets.*

*4.2.1 Current Usage of CODO*

The application of CODO in various studies demonstrates its versatility and significance in the domain of COVID-19 research. For annotating EHRs, CODO has been used along with other ontologies such as COVID-19 and COVIDCRFRAPID, and SNOMED CT (Keloth et al., 2020). CODO's concepts have been reused and aligned with external data sources in ontologies focusing on government responses to COVID-19 (Qundus et al., 2021). It has also been utilized in frameworks for systematic conflict resolution and bias assessment in ontologies, addressing issues like gender-specific socio-cultural biases and other biases such as socio-cultural biases (Keet, 2021; Keet & Grütter, 2021). Additionally, CODO has been instrumental in the development of new ontologies, such as the COVID-19 Diagnosis Ontology (CDO), which leverages existing knowledge bases for enhanced diagnostic capabilities (Wu et al., 2021), and in the COVID-19 Pandemic Ontology that integrates various COVID-19 and other domain ontologies to manage large volumes of pandemic-related data (González-Eras et al., 2022). The ontology has also been crucial in Chabot frameworks designed to deliver COVID-19 information effectively to older adults (Wang et al., 2021), and in developing models for conformance checking of decision mining rules, utilizing CODO for collecting and analyzing COVID-19 case and patient information (Toumia & Mejri, 2021). Furthermore, CODO's concepts have been used in the analysis of infection chains in smart cities, enhancing the understanding of COVID-19 spread and control measures (Silega et al., 2022) . In the iTelos project, CODO concepts were manually matched to enhance data shareability and reusability, further showcasing its application in building reusable knowledge graphs (Giunchiglia et al., 2021). The integration of CODO in these diverse applications highlights its essential role in advancing COVID-19 analytics, knowledge representation, and decision-making processes across multiple research initiatives (Lin et al., 2022).

## 5. Conclusion

CODO is a formal ontological framework specifically developed for the aggregation, representation, analysis, and visualization of diverse COVID-19 data. The latest version, CODO v1.5, introduces a comprehensive set of terms and relationships that capture a broad spectrum of COVID-19-related information, making it an indispensable tool for addressing the multifaceted challenges posed by the pandemic. CODO's modular structure, which encompasses 385 classes, 214 object properties, and 123 data properties, facilitates seamless integration and expansion, ensuring interoperability and enhancing data usability for various stakeholders, including medical professionals, hospitals, policymakers, government agencies, and application developers. CODO's structured approach and adherence to best practices and W3C standards ensure data integration and semantic interoperability. The ontology can support a wide range of applications, such as search engines, question-answering systems, risk detection tools, document annotation tools, data knowledge graphs, and analytics tools. These applications are crucial for effectively navigating and interpreting the vast amounts of COVID-19 data, thereby aiding in informed decision-making and efficient management of the pandemic. The development and continuous enhancement of CODO demonstrate the importance of ontological frameworks in managing complex data ecosystems. Future work will involve creating a manuscript detailing the integration and aggregation of various modules, as well as further enhancing CODO's capabilities. As the pandemic progresses, CODO will continue to evolve, adapting to new data and insights to remain an essential tool in ongoing efforts. With its comprehensive coverage and structured methodology, CODO exemplifies the crucial role of ontologies in artificial intelligence and data management during global health crises.